\documentclass[twocolumn,showpacs,preprintnumbers,amsmath,amssymb,aps]{revtex4}

\usepackage{graphicx}
\usepackage{dcolumn}
\usepackage{bm}

\begin{document}

\preprint{Physical Review Letters, in press}

\title{Heating of the Solar Corona by Dissipative Alfv\'en Solitons}
\author{K. Stasiewicz}
\affiliation{Swedish Institute of Space Physics,  SE-751 21 Uppsala,
Sweden} \altaffiliation[Also at ]{Space Research Center, Polish
Academy of Sciences, Warsaw, Poland.}

\email{k.stasiewicz@irfu.se} \homepage{www.cluster.irfu.se/ks/}


\begin{abstract}
Solar photospheric convection drives myriads of dissipative Alfv\'en
solitons (hereinafter called alfvenons) capable of accelerating
electrons and ions to energies of hundreds of keV and producing the
X-ray corona. Alfvenons are exact solutions of two-fluid equations
for a collisionless plasma and represent natural accelerators for
conversion of the electromagnetic energy flux driven by convective
flows into kinetic energy of charged particles in space and
astrophysical plasmas. Their properties have been experimentally
verified in the magnetosphere, where they accelerate auroral
electrons to tens of keV.
\end{abstract}


\maketitle

Understanding the mechanisms that heat plasma in the solar corona to
temperatures of millions of Kelvin  has been a long standing problem
in solar physics. The early view was that the convection in the
photosphere produces sound waves, internal gravity waves, and
magnetohydrodynamic waves which propagate upward to the solar corona
and deposit their energy to the ambient gas. Later, it was realized
that all but Alfv\'en waves are dissipated and/or refracted in the
chromosphere and in the transition region before reaching the corona
\citep{Withbroe:77,Kuperus:81,Priest:82,Zirin:88}.  The
electromagnetic energy driven by convective flows can be transported
along the magnetic field as Poynting  flux of Alfv\'en waves.
However, Alfv\'en waves are disinclined to dissipate in
collisionless plasmas, and the main problem was to explain how this
energy flux is deposited locally to heat particles in the solar
corona \citep{Parker:88}. During the last 50 years there have been
many attempts to solve this outstanding problem in astrophysics, and
there are more than 20 different models and mechanisms for coronal
heating proposed in the literature; see reviews
\citep{Narain:96,Mandrini:2000,Aschwanden:2004}.

In this Letter I show  that two-fluid equations for a collisionless
plasma have nonlinear solutions in the form of dissipative  Alfv\'en
solitons (alfvenons), which represent filamentary space charge
structures with strong perpendicular electric fields and large
parallel potential drops. Alfvenons will form spontaneously when a
magnetohydrodynamic perturbation propagates upward in the solar
corona where the Alfv\'en speed decreases. They represent natural
plasma accelerators for converting electromagnetic energy flux
driven by convective flows to kinetic energy of charged particles on
spatial scales related to the ion inertial length
$\lambda_i=c/\omega_{pi}$, where $c$ is the speed of light and
$\omega_{pi}$ is ion plasma frequency. Alfvenons can provide an
explanation for various aspects of electromagnetic energy
dissipation and heating in the solar corona and in the planetary
magnetospheres.

\begin{figure}
\includegraphics[width=0.9\columnwidth]{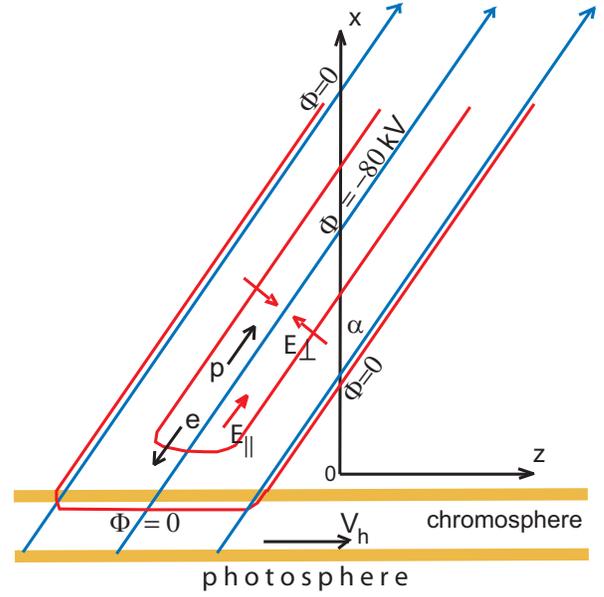} \caption{\label{Fig1} Geometry and coordinate system
for a wave structure (alfvenon) propagating along $x$ at angle
$\alpha$ to the magnetic field (blue lines), and driven by
convective flows in the photosphere. The electric potential
structure (red lines) would accelerate electrons toward the
chromosphere and ions out of the chromosphere. The angle
$\alpha=\sin^{-1}(V_h/V_A)$ is small.}
\end{figure}
Consider the geometry of magnetic fields in the solar corona
depicted in Fig.~\ref{Fig1}. It shows magnetic field lines (blue)
inclined at angle $\alpha$ from the wave propagation direction $x$.
There is a transverse  flow with speed $V_h$ in the $z$ direction,
associated with the electric field $\mathbf{E}=-\mathbf{V}_h\times
\mathbf{B}$. This geometry implies that the photospheric convection
drives electromagnetic energy flux
$\mathbf{S}=\mu_0^{-1}\mathbf{E}\times \mathbf{B}$  with an upward
component $ S_x=\mu_0^{-1} V_h B^2 \sin\alpha \cos\alpha$. Dragging
of magnetic field lines by variable convection of a conductive
plasma induces magnetic stresses propagating along \textbf{B} with
Alfv\'en speed $V_A=B(\mu_0 \rho_i)^{-1/2}$, which implies that
$\sin\alpha=\langle V_h \rangle/V_A$. The upward energy flux carried
by Alfv\'en waves driven by convection $\langle V_h \rangle$ is then
\begin{equation}\label{sx2}
    S_x=(\rho_i/\mu_0)^{1/2} \langle V_h \rangle^2 B \cos\alpha  .
\end{equation}
This form of the energy flux is consistent with an empirical
relation  between the total energy deposition and photospheric
magnetic flux in active regions, $S_x \propto B^\kappa$ with
$\kappa\approx 1$, while $\kappa\approx 2$ should be observed if the
energy dissipation was related to  reconnection of tangled magnetic
field lines \citep{Fisher:1998}. It is estimated that an average
upward energy flux of $ 10^7$ erg cm$^{-2}$ s$^{-1}$ is sufficient
to account for energy deposition in the solar corona
\citep{Withbroe:77}. For the magnetic field $B \approx 100$ G and
photospheric density of $\rho_i=10^{-7}$ g cm$^{-3}$, a convection
speed as small as $V_h=0.5$ km/s provides Poynting flux exceeding $
10^7$ erg cm$^{-2}$ s$^{-1}$ (10 kW m$^{-2}$). It is therefore
justified to assume that the electromagnetic energy flux of Alfv\'en
modes driven by the photospheric convection ($V_h>0.5$ km/s) is
sufficient to power all processes in the chromosphere and the solar
corona. It will be shown below that the dissipation mechanism for
this Poynting flux can be provided by nonlinear Alfv\'en structures
similar to those observed in association with auroral acceleration.

The height distributions of the background magnetic field $B(x)$ and
of ion density $\rho_i(x)$ in the solar atmosphere is such that the
Alfv\'en speed decreases with $x$ from a maximum value inside the
chromosphere \citep{Aschwanden:2004}. It is known that Alfv\'en
waves can become evanescent or nonlinear if they propagate in a
region of decreasing Alfv\'en speed \citep{Stasiewicz:jgr05}. One
expects then formation of nonlinear Alfv\'en waves (alfvenons) at
some altitude in the corona. From the general dispersion equation
for a two-fluid plasma model  one can determine the locations of
nonlinear waves in the phase space ($M, \alpha$), where
$M=V_{x0}/V_A=\omega/kV_A$ is the alfvenic Mach number.
\begin{figure}
\includegraphics{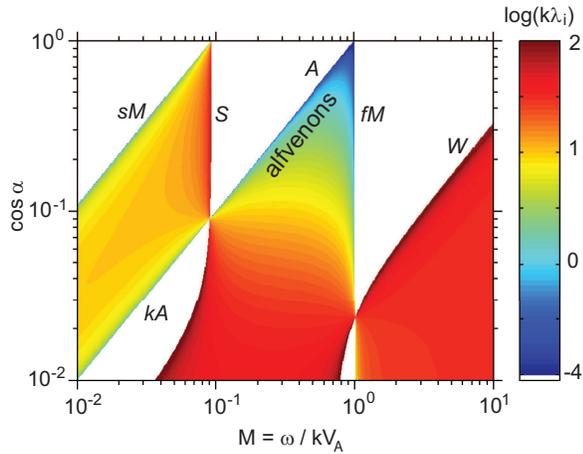}
\caption{\label{Fig2} Occurrence region of nonlinear waves in
two-fluid theory for plasma $\beta=10^{-2}$, with marked position of
alfvenons. Color coded is $\log (k\lambda_i)$, the exponential
growth rate of nonlinear waves. White regions are occupied by linear
waves.}
\end{figure}
Figure \ref{Fig2} shows portraits of the exponential spatial growth
rate $k\lambda_i$ of nonlinear waves \citep{Stasiewicz:jgr05}. White
areas in Fig.~\ref{Fig2} correspond to the well known linear waves:
slow magnetosonic ($sM$), sound ($S$), Alfv\'en ($A$), kinetic
Alfv\'en waves ($kA$), fast magnetosonic ($fM$), and whistler modes
($W$). The position of the sound line ($S$) is determined by the
sound speed $V_s=(\gamma\beta/2)^{1/2}V_A$, where $\gamma$ is the
polytropic pressure exponent and $\beta$ denotes the ratio of
plasma/magnetic field pressures. A linear Alfv\'en wave from region
$A$ in Fig.~\ref{Fig2} can tunnel to the nonlinear alfvenon region
(below the line $M=\omega/kV_A(x)=\cos\alpha$), if $V_A(x)$ is
decreasing along the wave path.

In a low beta plasma, details of the pressure model
\citep{Stasiewicz:PRL05} are not relevant and the electron inertia
is not important for modes propagating quasi-parallel. The governing
equations for a magnetohydrodynamic structure propagating along $x$
can be written in a stationary wave frame as
\citep{Stasiewicz:PRL04,Stasiewicz:jgr05}
\begin{eqnarray}\label{db1}
\frac{\lambda_i}{M_\alpha}\frac{\partial b_y}{\partial x} =
b_{z0}(n-1)-b_{z}\left(1-\frac{n}{M_\alpha^{2}}\right), \\
\label{db2} \frac{\lambda_i}{M_\alpha}\frac{\partial b_z}{\partial
x}=b_y\left(1-\frac{n}{M_\alpha^{2}}\right),\\
\label{db3} \frac{\partial n}{\partial x}
=\left(\frac{M_0^2}{n^{2}}-\frac{\gamma\beta}{
2}n^{\gamma-1}\right)^{-1}
\left[b_y\frac{\partial b_y}{\partial x}+ (b_{z0}+b_z)\frac{\partial b_z}{\partial x}\right],\\
\label{db4} e_x=-M_0 b_y \tan\alpha
-\frac{\gamma\beta_e}{2}\lambda_i \frac{\partial n^\gamma}{\partial
x},
\end{eqnarray}
with $M_\alpha=M_0/\cos\alpha, \; \beta=2\mu_0 p_{
0}/B_0^2,\,b_{z0}=\sin\alpha +g(x)$, where $g(x)$ represents a weak
gradient of the background magnetic field, which can be used to
study the transition between the linear and nonlinear regimes.
Subscript '0' denotes background quantities at the starting point
$x_0$. The magnetic field is normalized with $B_0$, the electric
field $e_x=E_x/V_AB_0$, and $n=N/N_0$ is the ion number density.
Equations (\ref{db1})-(\ref{db4}) represent a complete set of fully
nonlinear Hall-MHD equations for field variables $b_y,b_z,n,e_x$.
Note that $b_x=\cos\alpha,\; e_y=M_0\sin\alpha,\; e_z=0$ are
constant, $g(x_0)=0$, and the variations of the background density
due to gravitation are neglected.

The above equations describe linear (sinusoidal) as well as
nonlinear (cnoidal) waves, including solitons, in branches:
slow/fast magnetosonic, Alfv\'en, acoustic, and kinetic Alfv\'en
waves. They are easily integrated with an initial perturbation
$\delta b_z$, implying the boundary values:
$b_y(x_0)=0,\,b_z(x_0)=\delta b_z,\,
n(x_0)=1+(M_0^2-\gamma\beta/2)^{-1}b_{z0}\delta b_z$.
\begin{figure}
\includegraphics{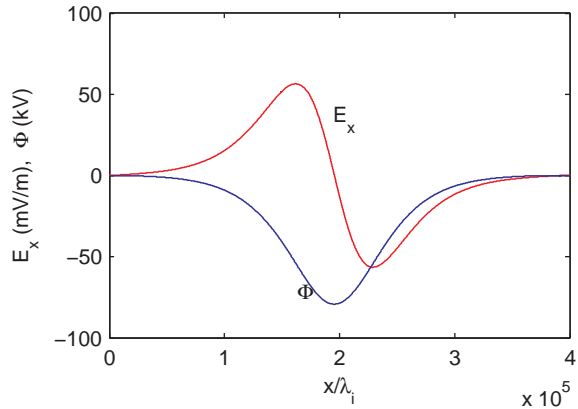}
\caption{\label{Fig3} Exact solutions of  equations
(\ref{db1}-\ref{db4}) for an alfvenon propagating at angle
$\alpha=0.5^\circ$. Medium parameters: $B_0=10$ G, $N_0= 10^8$
cm$^{-3},\; \beta=10^{-2},\; \gamma=1.6$, and $\lambda_i=20$ m.}
\end{figure}
Figure \ref{Fig3} shows exact solutions for the electric field
$E_x$, and the electric potential $\Phi=-\int E_x dx$.  The magnetic
signatures and field-aligned current associated with this structure
are shown in Fig.~\ref{Fig4}.  The structure  has an oppositely
directed electric field ($\nabla \cdot \mathbf{E}<0$) with a
negative potential in the center, i.e. it is an ion hole. The total
magnetic field has a small depression correlated with the density
depression.
\begin{figure}
\includegraphics{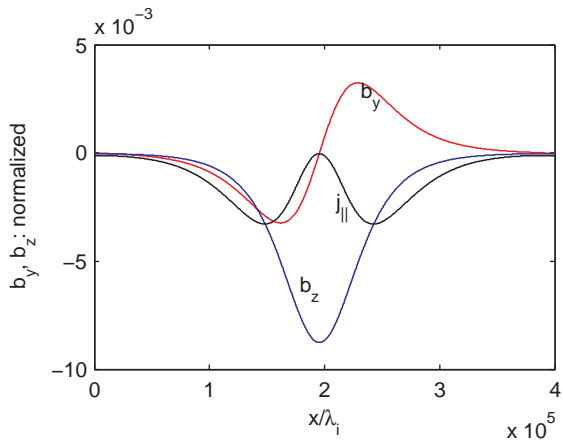}
\caption{\label{Fig4} Field-aligned current density $j_\parallel$
(arbitrary units, $j_{min}=-2\times 10^{-8}$ A m$^{-2}$), and
transverse magnetic field perturbations ($b_y,b_z$, normalized with
$B_0$) for the alfvenon in Fig.~\ref{Fig3}. }
\end{figure}
The computations were made in dimensionless variables and converted
to physical units using typical values measured in the corona:
$B\approx 10$ G and $N\approx  10^8$ cm$^{-3}$, which imply
$V_A\approx 2000$ km/s, and $\lambda_i\approx 20$ m. The computed
electric field can be schematically presented as equipotential (red)
lines shown in Fig.~\ref{Fig1}. It is seen that the magnetic field
lines external to the structure have potential equal to zero. The
chromospheric potential is also zero because it is along the
convection streamlines. However, the center magnetic field line is
at a large negative potential (--80 kV in Fig.~3), implying the
existence of a parallel potential drop equal to the perpendicular
potential drop. Conservation of total energy of a particle (kinetic
+ electric potential) implies that an electron leaving the structure
along the magnetic field lines to the chromosphere will acquire
kinetic energy equivalent to $e\Phi(x)$, producing X-rays when
interacting with the ambient gas. Similarly, an ion from the
chromosphere entering the structure would be accelerated upward by
the same potential $\Phi(x)$. Mathematically, this nonlinear
structure is formed as a result of a balance between nonlinear
growth and dispersion related to terms with ion inertia $\lambda_i$
in the governing equations (\ref{db1})-(\ref{db4}), and therefore it
could be regarded as a soliton. However, connection with the
conducting chromosphere introduces a dissipative element to this
soliton. Its energy would dissipate by acceleration of electrons
toward the chromosphere and ions out of the chromosphere. Alfvenons
form filamentary structures with extension along the magnetic field
much larger than the perpendicular size. In the example of Fig.~3,
the length of the alfvenon is $L\sim 3\times 10^5\,\lambda_i\approx
6000$ km, while the width $L_\perp\approx L\sin\alpha\approx 50$ km.
The length of the alfvenon would increase (decrease) with decreasing
(increasing) angle $\alpha=\sin^{-1}(V_h/V_A)$.

Let us estimate the  electric field available for acceleration of
charged particles in alfvenons. From Eq.~(\ref{db4}) we find that $
E_x \approx - M V_A B_y\tan\alpha$. A typical magnetic polarization
pattern for alfvenons, which can be found from numerical solutions,
is rotation of the transverse magnetic field ($B_y,B_z$) around the
guiding $B_x$ field such that $|B_y|_{max}\sim
B_{z0}=B_0\sin\alpha$. Furthermore, alfvenons are formed near $M
\approx \cos\alpha$ (see Fig.~\ref{Fig2}), which implies
\begin{equation}\label{Ex}
    |E_x|_{max}\sim  V_A B\sin^2\alpha.
\end{equation}
As is seen, the electric field in an alfvenon depends on the
propagation angle and can vary from zero (parallel propagation) to
the maximum value of $E_A=B_0 V_A$ for perpendicular propagation.

The electric potential structure obtained here in a self-consistent
way (though in a simplified geometry) is not a speculative result
awaiting observational verification. In fact, the presence of such
electric potential structures above the aurora was first inferred
from satellite measurements 30 years ago \citep{Gurnett:72}.
Properties of these U-shaped auroral acceleration structures have
been thoroughly investigated in numerous publications listed in a
recent review of auroral processes \citep{issi:03}, but until now
there was no theoretical model for these structures. Thus, the
alfvenons described in this Letter provide also an explanation for
the U-shaped auroral potential structures.  Particle measurements
inside these acceleration structures show upward directed ion beams
associated with electrons accelerated downward by the electric
potentials 1-20 kV, and forming so-called inverted-V structures in
energy versus time flux-spectrograms.  An obvious difference between
the solar corona and the magnetosphere is that in the first case the
driving convection is applied to the bottom end of the flux tubes in
Fig.~\ref{Fig1} (in the photosphere), while in the second case to
the top end (in the distant magnetosphere). The magnetospheric
alfvenons are created below an altitude of 1 R$_E$, i.e. in region
where the Alfv\'en speed starts to decrease toward the ionosphere,
as predicted by this theory. Similar acceleration structures have
been recently measured on Mars \citep{Lundin:06}, indicating the
universal applicability of the alfvenon mechanism. One should also
mention that the present theory of nonlinear waves, when extended
with anisotropic and polybaric pressure equations
\cite{Stasiewicz:PRL05}, gives numerical results in quantitative
agreement with observations of large amplitude ($\delta B/B\sim
200\%$) trains of magnetosonic solitons in a hot ($\beta \sim 10$)
magnetosheath plasma \citep{Stasiewicz:grl04,Stasiewicz:jgr05}.
Plasmas in the solar atmosphere and the magnetosphere are quite
similar, with the same range of plasma beta ($10^{-3}-10^1$), and a
similar altitude transition from a partly ionized, collision
dominated to a fully ionized, collisionless plasma.

The numerical results obtained in this paper indicate that depending
on the propagation angle and parameters $V_A,\,B,\,\beta$ in
different regions of the corona one could possibly create
acceleration structures with electric potential drops that would
account for most of the observed X-ray and radio emissions due to
accelerated electrons. Typical integrated voltages across the
alfvenon structures in the solar corona determined numerically are
hundreds of kilovolt. Because the photospheric convection is a
variable and permanently occurring process, myriads of alfvenons
will be continuously recreated at different altitudes over the whole
Sun, producing what is observed as the X-ray corona. Alfvenons could
form bundles of threads or sheets and build up larger current
structures, which should be seen in X-ray emissions when accelerated
electrons thermalize in denser regions. Actually, high-resolution
images from TRACE (Transition Region and Coronal Explorer) are
suggestive of the existence of such threads, sheets and arcades of
energized particles in the solar corona.

The present theory provides a natural link to explanations for the
acceleration of the solar wind and the creation of solar flares. The
solar wind originates from the coronal regions cool in X-rays, which
have open magnetic field lines. Solar wind ions have energy
concentrated in the drift velocity with a small thermal spread
($V_d\gg V_t$), which is consistent with acceleration by electric
potential difference, and not by any stochastic or turbulent process
that would produce $V_d \sim V_t$. As is seen in Fig.~\ref{Fig1} an
elementary alfvenon event ejects ions outward, providing the first
kick-off for the solar wind.  Typical speeds of the solar wind
protons 400--800 km/s could be produced by an electric potential of
1--4 kV. A possible explanation for such small values of the
acceleration voltage in coronal holes is that plasma on open field
lines cannot support significant parallel electric fields.

On closed coronal loops, periodic energy releases by alfvenons
driven by photospheric convection would heat plasma that will remain
confined in magnetic traps. This would allow large pressure
gradients and pitch-angle anisotropy to build up and support
increasingly larger parallel electric fields that would accelerate
ions upward. When the magnetic field tension at the top of the loop
will not be able to balance the Reynolds stresses associated with
flows of accelerated ions,  the loop would break up, leading to
plasma jets, coronal mass ejections, and a topological
reconfiguration of the magnetic field.  A detailed description of
the mechanism for flares implied by the present theory is beyond the
scope of this Letter and will be addressed elsewhere.

Application of the present model to heating of the solar corona can
be summarized as follows:

(a) Dragging of magnetic field lines by non-stationary photospheric
convection $V_h$ induces magnetohydrodynamic perturbations
propagating upward  and carrying energy flux given by
Eq.~(\ref{sx2}).

(b) An upward propagating magnetic perturbation becomes nonlinear in
regions of decreasing $V_A(x)$ and forms an alfvenon, i.e. the
U-shaped electric potential structure shown in Fig.~\ref{Fig1}. The
size of the structures scales with $\lambda_i$ and the propagation
angle. For a propagation angle $\alpha\approx 0.5^\circ$ and
$\lambda_i\sim 20$ m the size is: $L_\parallel\sim 6000$ km,
$L_\perp \sim 50$ km, and the generated voltage is $\Phi\sim 100$
kV.

(c) The electric potential structures created in the solar corona
dissipate electromagnetic energy through direct acceleration of
electrons toward the chromosphere and ions out of the chromosphere.

(d) Ions accelerated outward by 1--4 kV potential drops in coronal
holes would initiate outflow of the solar wind.

(e) Electrons accelerated toward the chromosphere by a potential
difference up to hundreds kV (on closed loops) produce X-ray corona,
radio emissions, and evaporation/heating of the chromosphere.

(f) The processes described in (a)-(e) would repeat as long as there
is variable photospheric convection pumping the energy flux (1) into
regions of a decreasing Alfv\'en speed.

The alfvenons introduced in this Letter appear to be effective and
spectacular converters of electromagnetic energy flux into kinetic
energy of particles. They have been measured by numerous spacecraft
in  the terrestrial magnetosphere, where they accelerate auroral
electrons toward the ionosphere to  tens of keV, and are detected
also in the martian environment. They appear to be of universal
importance for astrophysical plasmas and must occur also in the
solar atmosphere, where they can account for the acceleration and
heating of plasma in the solar corona.

\end{document}